# Neutron optical tuning of Fe/$^{11}$B$_4$CTi multilayers for optimal polarization and increased reflectivity for polarizing neutron optics

A. Zubayer,*[1] N. Ghafoor,[1] K.A. Thorarinsdottir,[2] A. Glavic,[3] J. Stahn,[3] G. Nagy,[4] A. Vorobiev,[4] F. Magnus,[2] J. Birch,[1] F. Eriksson[1]

1. Department of Physics, Chemistry and Biology (IFM), Linköping University, SE-581 83 Linköping, Sweden

2. Science Institute, University of Iceland, Dunhaga 3, IS-107 Reykjavik, Iceland

3. Paul Scherrer Institut, 5232, Villigen, Switzerland

4. Department of Physics and Astronomy, Uppsala University, SE-75120, Uppsala, Sweden

* anton.zubayer@liu.se, +46760583232

## Abstract

**The concept of scattering length density tuning for improved polarization is investigated for Fe/$^{11}$B$_4$CTi multilayers and compared to the commonly used Fe/Si system in polarizing multilayer neutron optics. X-ray and neutron reflectivity, magnetization, and neutron polarization have been measured on such multilayers, highlighting differences from conventional Fe/Si multilayers. The multilayer systems were deposited with 25 Å period thickness, a layer thickness ratio of 0.35 and 20 periods, using ion-assisted DC magnetron sputtering. Replacing Si with $^{11}$B$_4$CTi for these multilayers showed an increase in reflectivity due to a reduction in interface width. By tuning the ratio between $^{11}$B$_4$C and Ti in the non-magnetic layers, a broad range of scattering length density values was achieved, facilitating scattering length density contrast matching between layers for spin-down neutrons,** **thereby enhancing polarization. These findings demonstrate the potential of Fe/$^{11}$B$_4$CTi multilayers as a promising option for polarizing neutron optics and highlight the concept of scattering length density tuning in a large range using $^{11}$B$_4$CTi.**

**Introduction**

Polarized neutrons play a crucial role in investigating magnetism, spintronics, and biological samples, as they offer distinct contrast based on their spin states (here referred to as spin up and spin down) [1]. Multilayer neutron optics are essential components for guiding, collimating, polarizing, and analysing neutrons in neutron scattering facilities, spanning from the source to the detector [2]. These optics consists of repetitive layers, either periodic or depth-graded, designed for single wavelength or broadband reflection/polarization, respectively. Optimal neutron reflectivity and polarization characteristics require a scattering length density (SLD) contrast when selecting the material system for multilayer optics. While non-polarizing optics rely on a large difference in nuclear scattering length densities, typically using Ni and Ti, polarizing optics utilize layers of ferromagnetic (e.g., Fe, Co) and non-magnetic materials (e.g., Si, Ti) to maximize contrast in scattering length density for one spin state while minimizing it for the opposite spin [1],[3]. The scattering length density comprises both a nuclear component (n-SLD) and a magnetic component (m-SLD), with the nuclear part determined by the elements' neutron scattering cross section and the layers' mass density, while the magnetic part is influenced by the magnetic moment [1],[4]. To achieve maximum polarization, the following relations needs to be satisfied;

**Spin up:** $\rho_{(n,FM)} + \rho_{(m,FM)} \gg \rho_{(n,NM)}$

**Spin down:** $\rho_{(n,FM)} - \rho_{(m,FM)} = \rho_{(n,NM)}$

where $\rho$ is the SLD, m for magnetic, n for nuclear, FM for ferromagnetic layer and NM for non-magnetic layer.

An ideal multilayer has flat and sharp interfaces, but in reality interface imperfections such as roughness, intermixing and magnetically dampened silicide layers hamper the reflectivity, polarization and possibility to reach higher supermirror m-values, where the m-value represent the critical angle of the supermirror compared to that of Ni. State-of-the-art Fe/Si neutron optics encounter difficulties in achieving effective reflectivity and polarization at high angles due to the interface roughness and SLD mismatch. These challenges arise primarily from the formation of iron silicide at the interfaces, resulting from the high reactivity of Si with Fe. The presence of iron silicide leads to deviations in both nuclear and magnetic SLD, as well as significant roughness caused by crystalline facets at the interfaces due to the polycrystalline microstructure of Fe and/or iron silicide. Attempts to circumvent the silicide issue by substituting Si with Cr or Zr have proven unsuccessful, as these elements tend to alloy with Fe, thus hindering the desired outcomes in a similar manner [5],[6],[7]. Consequently, the development of well-performing polarizing neutron optics with thin periods remains a challenge [8],[9],[10],[11]. Achieving thinner periods is necessary to enable reflectivity at higher angles, while simultaneously maintaining SLD contrast and matching to attain maximum polarization.

Addressing the problems associated with iron silicide formation, interface roughness, and SLD contrast will facilitate higher reflectivities and more efficient polarization, even at larger scattering angles.

This study aims to replace Si layers with $^{11}B_4CTi$ layers in order to address both interface structure and SLD considerations associated with the nonmagnetic layer. By replacing Si, the iron silicide formation at the interfaces is avoided. Furthermore, the inclusion of $B_4C$ in the replacement layer, known for its ability to amorphize and smoothen interfaces, offers the potential for improved reflectivity [12],[13],[14],[15]. It is important to note that natural B consists of 20% high neutron absorbing $^{10}B$ and 80% low neutron absorbing $^{11}B$ isotopes, which is why isotope-enriched $^{11}B_4C$ is employed as part of the nonmagnetic layer to replace Si [16]. However, Fe/$^{11}B_4C$ multilayers are not suitable for achieving high neutron reflectivity due to the significantly high SLD of $^{11}B_4C$ ($9.011 \cdot 10^{-6}$ Å$^{-2}$)[17],[18]. Therefore, to compensate for the high SLD of $^{11}B_4C$ it is proposed in this study to mix it with an element having a negative SLD, such as Mn or Ti, which are the elements with the lowest SLD values. Ti, which is already extensively studied and used in neutron optics, becomes the preferred choice for this purpose [19],[20]. Consequently, Fe is paired with an alloy of $(^{11}B_4C)_xTi_{1-x}$ to create the desired material composition. For convenience $(^{11}B_4C)_xTi_{1-x}$ will be written as $^{11}B_4CTi$ even though there are compositional deviations from stoichiometric $^{11}B_4CTi$. This denotation is simply the combination of using $^{11}B_4C$ and Ti as sputter targets during deposition. A previous study on SLD tuning using a mix of Si and $^{11}B_4C$ has been performed, however such a mix could only yield an SLD higher than Si and not lower [21].

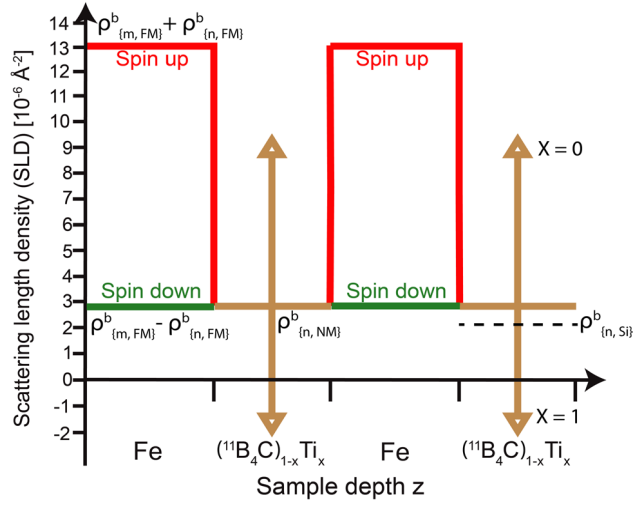

**Figure 1.** Schematic illustration of the SLD depth profile of a Fe/$^{11}$B$_4$CTi multilayer for spin up (red) and spin down (green) neutrons. The arrows indicate the range of possible SLD tuning of the non-magnetic layers by varying the ratio between $^{11}$B$_4$C and Ti.

Figure 1 illustrates the tuneability of the SLD to achieve optimum polarization by matching the SLD of $^{11}$B$_4$CTi with the spin-down SLD of Fe. Example simulations can be seen in Figure S4. This tuneability provides the flexibility to adjust the SLD regardless of the magnetic SLD of the magnetic layer, which can deviate from bulk values due to factors such as the thickness of the magnetic layer, layer thickness ratio, and magnetic moment.

**Results and Discussion**

Fe/Si, Fe/$^{11}$B$_4$C, and Fe/$^{11}$B$_4$CTi multilayer samples were grown on Si substrates using ion-assisted DC magnetron sputter deposition. The multilayers have a nominal bilayer thickness of $\Lambda = 25$ Å, a nominal layer thickness ratio of $\Gamma_1 = 0.35$, where $\Gamma_1$ is the Fe layer thickness to bilayer thickness ratio, and N = 20 bilayers. $\Gamma_1 = 0.35$ was chosen since a layer thickness ratio $\Gamma_1$ close to 0.5 gives the highest reflectivity for the first Bragg peak, however, would result in the elimination of every even number ordered Bragg peak. Since fitting the second order Bragg peak is crucial for a reliable fit, a layer thickness ratio further away from 0.5 was needed. Hence $\Gamma_1 = 0.35$. It is important to note that for Fe/Si multilayers the obtained layer thickness ratio deviated from this nominal value due to intermixing between the Fe and Si layers, possibly due to silicide formation. However, it is also important to note that regardless of the thickness of the magnetic layer, the amount of Fe atoms between each sample is kept the same since the growth conditions and time to sputter Fe were identical.

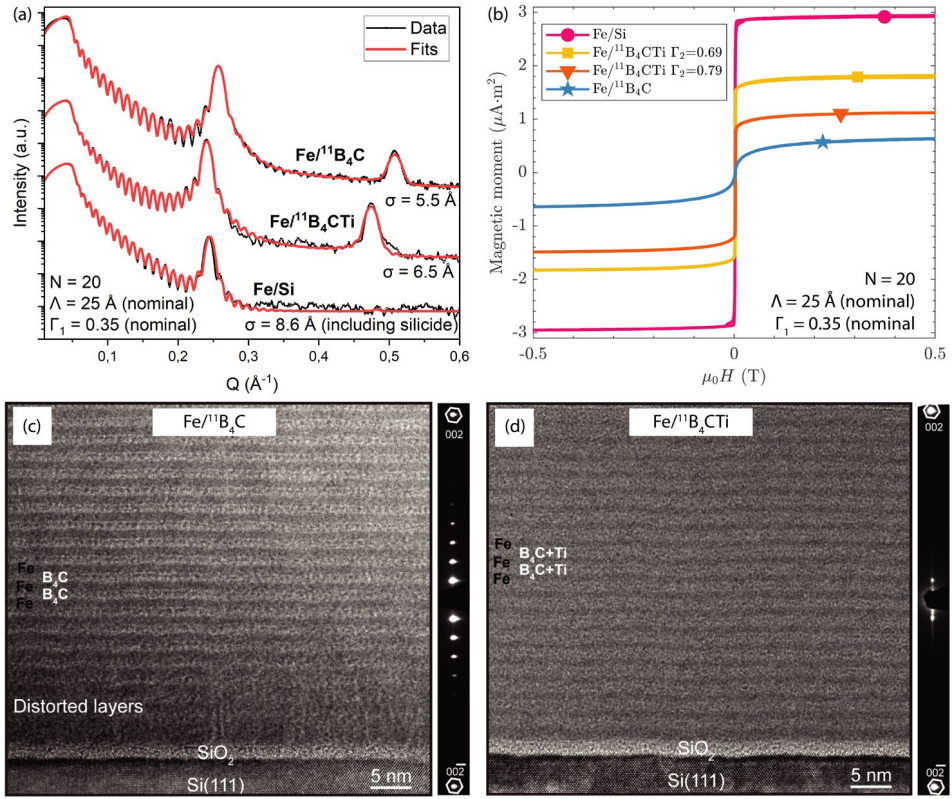

**Figure 2.** (a) X-ray reflectivity (XRR) measurements (black) and fits (red) of Fe/$^{11}$B$_4$C, Fe/$^{11}$B$_4$CTi and Fe/Si. All samples were aimed to have a bilayer thickness of 25 Å and 20 bilayers. Through fitting using the Parratt formalism in GenX [22] we obtained the interface width σ. (b)

Vibrating sample magnetometry hysteresis loops of Fe/Si (nominal), Fe/$^{11}$B$_4$C, and Fe/($^{11}$B$_4$C)$_x$Ti$_{1-x}$ with two different atomic percent ratios $\Gamma_2$ = 0.69 and 0.79, using vibrating sample magnetometry (VSM) at room temperature. All the samples had a bilayer thickness of around $\Lambda$ = 25 Å with N = 20 periods and a nominal layer thickness ratio of $\Gamma_1$ = 0.35. (c-d) Cross-sectional views of Fe/$^{11}$B$_4$C and Fe/$^{11}$B$_4$CTi as observed by transmission electron microscopy (TEM), with corresponding SAED. Both with a bilayer thickness of around $\Lambda$ = 25 Å with N = 20 periods and a nominal layer thickness ratio of $\Gamma_1$ = 0.35.

Figure 2 (a) presents X-ray reflectivity measurements and fits for Fe/$^{11}$B$_4$C, Fe/$^{11}$B$_4$CTi, and Fe/Si multilayers. It is observed that both Fe/$^{11}$B$_4$C and Fe/$^{11}$B$_4$CTi exhibit a more pronounced first-order Bragg peak compared to Fe/Si. Additionally, a distinct second-order Bragg peak is visible in Fe/$^{11}$B$_4$C and Fe/$^{11}$B$_4$CTi, while it is absent in the Fe/Si multilayer. The SLD profiles are shown in Figure S3. Figure 2(b) shows the magnetization curves of Fe/Si, Fe/$^{11}$B$_4$C, and Fe/$^{11}$B$_4$Ti with two different concentration ratios. The Fe/Si multilayers display the highest saturated magnetization, whereas Fe/$^{11}$B$_4$C exhibits the lowest. Modifying the concentration ratio, $\Gamma_2$, where $\Gamma_2$ denotes the atomic percent ratio of $^{11}$B + C in the non-magnetic layer, to increase the Ti content in the non-magnetic layer appears to enhance the saturated magnetization. Cross-sectional TEM micrographs in Figure 2 (c,d) compare the Fe/$^{11}$B$_4$C and Fe/$^{11}$B$_4$CTi multilayers.

The X-ray reflectivity (XRR) fits provided information about the bilayer thicknesses, and roughness and intermixing in the Fe/$^{11}$B$_4$C, Fe/$^{11}$B$_4$CTi, and Fe/Si multilayers. The bilayer thicknesses were found to be 24.8, 26.6, and 26.2 Å, respectively. The Fe/$^{11}$B$_4$C multilayer exhibited an interface width of 5.5 Å, and by adding Ti to a concentration ratio of $\Gamma_2$ = 0.74 in the non-magnetic layer, increased the interface width by only 1 Å to 6.5

Å. In contrast, the Fe/Si sample displayed large intermixing between Fe and Si, possibly forming iron silicide and roughness from the Fe layer. The fitting yielded an 8.6 Å interface width Fe. Since only one Bragg peak is present in the reflectivity measurement, a conclusive fit is however difficult to achieve, and the 8.6 Å interface width layer is considered to be a combination of roughness, intermixing and iron silicide, although the relative proportions are not possible to determine. The silicide formation, which leads to a change in layer thickness ratio, might contribute to destructive interference and the absence of the second order Bragg peak. The iron silicide formation in Fe/Si multilayers grown using DC magnetron sputter deposition are typically reported in the range 5.2 - 8.2 Å at each interface [8],[23], which is in good agreement with our results.

The second-order Bragg peak in Fe/$^{11}$B$_4$CTi was higher than in Fe/$^{11}$B$_4$C. This may appear counterintuitive since Fe/$^{11}$B$_4$C had a smaller interface width compared to Fe/$^{11}$B$_4$CTi. However, this can be explained by the X-ray scattering length density values of Fe, $^{11}$B$_4$C and $^{11}$B$_4$CTi, which are approximately 19.0, 20.2, and approximately 27.8·10$^{-6}$ Å$^{-2}$, respectively. The addition of Ti significantly increased the X-ray SLD contrast, leading to higher reflectivity despite the slightly larger interface width. The higher SLD contrast can also be seen in Figure S3.

In Fe/Si multilayers, the presence of an iron silicide layer at the interfaces should be considered. Therefore, when aiming for a layer thickness ratio of $\Gamma_1 = 0.35$ with a bilayer thickness of $\Lambda = 25$ Å, the nominal Fe thickness should be around 8-9 Å, while the Si layer should be around 16-17 Å. However, as mentioned previously, the thickness ratio $\Gamma_1$ was closer to 0.5 [8]. From a magnetic perspective, as seen in S1, it is known that multilayers with bilayer thicknesses up to approximately 70 Å can be influenced by interlayer exchange coupling (IEC) and finite size effects. In this thickness range the two competing effects influences the total magnetization. The IEC is the coupling through the non-magnetic layer which can increase the magnetization [24], while the finite size effect is the suppression of magnetization of the magnetic layer when it is near a non-magnetic layer [25]. Hence, in our observations the magnetization is indirectly affected by the magnetic layer thickness [26] and layer thickness ratio in ferromagnetic-based multilayers. The varying magnetization by the magnetic moment strongly impact the m-SLD, making pure Si unsuitable as a non-magnetic layer due to its SLD mismatch with spin-down Fe and its non-adjustable SLD. By having a tuneable SLD for the non-magnetic layer, it becomes possible to match the SLDs regardless of the m-SLD of the magnetic layer.

Figure 2 (b) shows that the Fe/Si multilayer had the highest magnetization, despite all multilayers having the same amount of Fe deposited in each film. This is due to the stronger IEC and weaker finite size effect which could be due to Fe being present throughout the entire multilayer even in the non-magnetic silicide layer. Furthermore, from the fits shown in Figure 2 (a) it was observed that the nominal layer thickness ratio was different for Fe/Si, meaning that the non-magnetic layer in Fe/Si was thinner than the non-magnetic layer in the Fe/$^{11}B_4C$ and Fe/$^{11}B_4C$ti multilayers [23],[27]. The increase in magnetization observed at higher concentrations of Ti and lower concentrations of $^{11}B_4C$ in Fe/$^{11}B_4C$Ti can be attributed to the aforementioned reasons, namely the IEC and finite size effect. In other words, an increase in the Ti ratio increased the IEC and/or decreased the finite size effect. Determining which of these effects that has the largest influence on magnetization is difficult, however not required since our proof of concept is the possibility to match the m-SLD regardless of the saturated magnetization [28]. From the TEM micrographs in Figure 2 (c,d), it is observed that the Fe/$^{11}B_4C$ multilayer has similar layer thicknesses compared to Fe/$^{11}$B$_4$CTi, which suggests that the increase in magnetization with higher Ti content is not due to a thinner non-magnetic layer, but dependent on the actual composition of the non-magnetic layer.

To accurately determine the atomic percentages of $^{11}$B$_4$C and Ti in the non-magnetic layer, time-of-flight elastic recoil detection analysis (ToF-ERDA) was employed. These values were utilized to generate the plot shown in Figure 3(a).

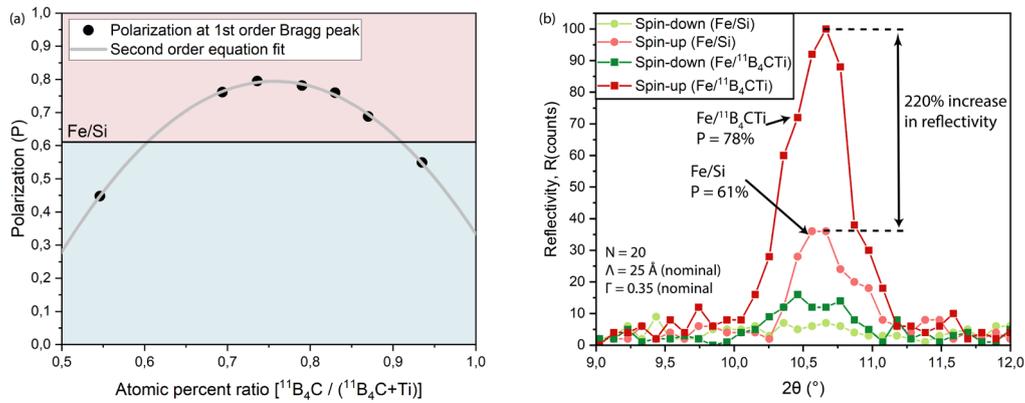

**Figure 3.** (a) Polarization of the first order Bragg peak for the Fe/$^{11}$B$_4$CTi samples containing different ratios of Ti and $^{11}$B$_4$C over a function of atomic percentage ratio of $^{11}$B$_4$C in the non-magnetic layer. The horizontal black line is the Fe/Si samples' polarization. The grey second-order polynomial fit is a guide for the eye for the expected polarization for various ratios of $^{11}$B$_4$C vs Ti. (b) Polarized neutron reflectivity between 9° to 12° for Fe/Si and the best polarizing Fe/$^{11}$B$_4$CTi sample.

The neutron polarization presented in Figure 3 (a) was calculated from the measured reflected intensities at the first order Bragg peak for both spin states of seven Fe/$^{11}$B$_4$CTi samples, as well as a Fe/Si sample. The atomic percentage ratio of $^{11}$B$_4$C in the non-magnetic layers was plotted against the polarization values, with black data points representing the Fe/$^{11}$B$_4$CTi samples and a horizontal line representing the polarization from the Fe/Si multilayer, serving as a reference value.

The tuneability of the polarization is evident in Figure 3 (a), and the local maxima correspond to the best polarization values. By adjusting the atomic percentage of $^{11}$B$_4$C in the non-magnetic layers, the polarization can be optimized. Figure 3 (b) displays the polarized neutron reflectivity measurements of Fe/Si and the best polarizing Fe/$^{11}$B$_4$CTi sample over the range of 9° to 12°, which covers the entire first-order Bragg peak. The critical edge region for the measurements can be seen in Figure S5. The flipping ratio, FR, which is the spin up reflectivity divided by the spin down reflectivity at the Bragg peak, for Fe/Si and Fe/$^{11}$B$_4$CTi are 4.2 and 8.8, respectively, corresponding to polarization values of 61% and 78%. Furthermore, the reflectivity for the spin up state in Fe/$^{11}$B$_4$CTi is 3.2 times higher than that of Fe/Si. The FR is improved by more than double, corresponding to a 17% increase in polarization. Based on the achieved polarization and reflectivity at this particular angle, it is possible to achieve polarizing supermirrors with m-values larger than 10, where commercial state-of-the-art polarizing neutron optics today only can reach m = 5.5 [29]. Meaning that if the Fe/$^{11}$B$_4$CTi material system was used with a depth graded period thickness going from thicker to as thin as 25 Å in period thickness, this could be achieved if Fe/$^{11}$B$_4$CTi was used with careful optimization of $^{11}$B$_4$C vs Ti ratio.

In this study, the aim was to overcome the limitations of Fe/Si multilayers in polarizing neutron optics by introducing Fe/$^{11}$B$_4$CTi as an alternative to Si. The Fe/$^{11}$B$_4$CTi multilayer system offers the potential to fine-tune the scattering length density and take advantage of the smoothening effects of $^{11}$B$_4$C, while avoiding formation of iron silicide layers. Vibrating sample magnetometry (VSM) measurements of Fe/$^{11}$B$_4$C revealed that the magnetization was affected by interlayer exchange coupling and finite size effects. Hence, the m-SLD of Fe would also vary with thickness and thickness ratio differences, emphasizing the importance of SLD tuning in to the non-magnetic layer to achieve high polarization.

TEM, XRR and VSM investigations showed that the Fe/Si multilayers with a nominal bilayer thickness of 25 Å deviated significantly from the intended thickness ratio between the magnetic and non-magnetic layers.

The incorporation of Ti into the non-magnetic layer in Fe/$B_4C$ multilayers was found to increase magnetization. By adjusting the ratio between $^{11}B_4C$ and Ti, the SLD of the non-magnetic layer could be fine-tuned to match the SLD of the Fe spin-down state, leading to optimal polarization. This was demonstrated through polarized neutron reflectivity, which also revealed that Fe/$^{11}B_4C$Ti outperformed Fe/Si in terms of reflectivity, suggesting its potential for reflectivity and polarization improvements at higher scattering vector values.

## Conclusions

The ability to adjust the SLD by using different ratios of $^{11}B_4C$ and Ti in the non-magnetic layer opens up opportunities for fine-tuning the SLD in various thin-film neutron applications. It would be interesting to further investigate the incorporation of $^{11}B_4C$ within the Fe layers to explore its smoothening effect and determine if iron silicide formation can be prevented by incorporating $^{11}B_4C$ in both Fe and Si layers.

Although the optimal concentrations of $^{11}B_4C$ and Ti for achieving polarization over a larger bandwidth need to be determined through extensive trials and simulations, the use of $^{11}$B$_4$CTi as a material to tune the SLD opens up new possibilities for advancing polarizing neutron optics and thin film neutron reflectivity research.

## Acknowledgments

This work was supported by the Swedish Research Council (VR), grant number 2019-04837_VR. The authors are grateful to PSI for granting beamtime for using the neutron reflectometry beamline MORPHEUS. A.Z acknowledge the grant from the Clark and Karen Bright Endowed Scholarship Honoring Angus Macleod from the SVC Foundation and the grant 2022-D-03 from the Hans Werthén Foundation. Accelerator operation, in Uppsala, is supported by the Swedish Research Council VR-RFI (Contracts No. 2017-00646_9 and 2019-00191) and the Swedish Foundation for Strategic Research (Contract No. RIF14-0053).

## Experimental Details

Fe/$^{11}$B$_4$C and Fe/$^{11}$B$_4$CTi multilayer thin films were deposited using ion-assisted DC magnetron sputter deposition in a high vacuum system with a background pressure of about $5.6 \cdot 10^{-5}$ Pa. The $10 \cdot 10 \cdot 0.5$ mm$^3$ 001-oriented single crystalline Si substrates were successively cleaned using trichloroethylene, acetone, and isopropanol, in an ultrasonic bath for 5 minutes and blown dry using N$_2$. Argon (99.999%) was used as the sputtering gas at a pressure of 0.51 Pa (3.8 mTorr) and the substrate was kept at an ambient temperature. In order to attract a high flux of low energy ions to the sample during growth, a bias voltage of -30 V was applied to the substrate and a substrate coil current of 5 A was used for the deposition of all samples. The three magnetron sources Fe, $^{11}$B$_4$C, and Ti were running continuously during deposition with material fluxes controlled by computer-controlled shutters. The two 3" sputtering targets Fe and Ti has purities of 99.5%, while the 2" sputtering target was used for $^{11}$B$_4$C (RHP Technology, 99.8% chemical purity, >90% isotopic purity). The magnetron discharges were established with power-regulated power supplies with powers of 20 W, 50 W, 50-150 W and 9-35 W were applied for Fe, Si, $^{11}$B$_4$C and Ti respectively. Co-sputtering of $^{11}$B$_4$C and Ti allowed for alloying two target materials with a desired composition. The low deposition rates allowed for more accurate layer thicknesses.

X-ray reflectivity measurements were conducted using a Malvern Panalytical Empyrean diffractometer with Cu-K$_a$ radiation and a PIXcel detector. On the incident beam side, a Göbel mirror was used with a 0.5° divergence slit, and on the diffracted beam optics side a parallel beam collimator with a collimator slit of 0.27° was used. Multilayer periods were calculated from the Bragg peak positions. X-ray diffraction was performed using a Panalytical X'Pert diffractometer in Bragg-Brentano geometry with a 2θ scanning range of 20° to 90°.

X-ray reflectivity data were used to determine the multilayer period and individual layer thicknesses as well as the interface roughness. The GenX software [30] was used for simulations to find the optimum ratio of $^{11}B_4C$ and Ti in the non-magnetic layer, achieving a match in SLD between the magnetic and non-magnetic materials. Furthermore, GenX was utilized to fit data obtained from X-ray and neutron reflectivity measurements. The software employs the Parratt recursion formalism to recursively simulate and fit reflectivity data, taking into account each interface in order to calculate the total intensity detected.

The A series were deposited to investigate the interlayer exchange coupling and finite size effects, but also the reflective properties compared to the reference Fe/Si sample R. Another series of samples was designed with varying ratios of Ti to $^{11}B_4C$ in the non-magnetic layer (indicated as B in Table 1 in SI) to determine if a specific ratio optimizes neutron polarization. All samples were carefully designed to ensure equal layer thicknesses within each period. The B series, along with the reference, R, Fe/Si sample, were used in both X-ray reflectivity (XRR) and polarizing neutron reflectivity (PNR) measurements to study the neutron reflection and polarization properties.

The magnetic properties of the samples were analyzed at room temperature using vibrating sample magnetometry (VSM) and magneto-optical Kerr effect (MOKE) techniques in longitudinal geometry. VSM measurements were conducted within the range -20 mT to 20 mT to obtain information on the samples' saturated magnetization and magnetic coercivity. On the other hand, MOKE provided higher resolution for the applied magnetic field but could not determine the absolute value of the magnetization amplitude.

Time-of-flight elastic recoil detection analysis (ToF-ERDA) was employed to determine the elemental composition of selected samples [31]. ToF-ERDA is an ion-beam based analytical technique that provides depth-resolved and quantitative information about the samples' composition, unaffected by matrix effects or the need for reference standards. For the ToF-ERDA measurements, 36 MeV $I^{8+}$ beam was used with an incident angle of 67.5° relative to the surface normal of the sample. The ToF-telescope and the gas ionization chamber detector was positioned at a 45° angle relative to the incident beam direction. This configuration allowed for the measurement of energy and flight time of the recoiled particles, providing information about the samples' elemental composition. The Potku code was employed for data analysis [32]. It is important to note that the depth resolution of ToF-ERDA was insufficient to distinguish the individual layers within the samples, as their thicknesses were on the order of a few Ångström. Therefore, the simulated and calculated composition is an effective value.

Polarized neutron reflectometry (PNR) experiments were carried out using the MORPHEUS instrument at the Swiss Spallation and Neutron Source (SINQ) at the in Paul Scherrer Institut (PSI), in Switzerland. In these experiments, a polarized neutron beam was directed at the sample at small incidence angles (θ) and reflect at each interface in our sample before being detected by a He-3 detector. PNR is sensitive to the spin-dependent SLD of the sample, providing information about the magnetization profile. The two possible spin states result in two distinct curves in the PNR measurements, and Bragg peaks are observed due to constructive interference. The measurements were performed on samples in an external magnetic field of approximately 20 mT and at 0° to 15° 2θ, using a wavelength of 4.825 Å.

Cross-sectional specimens for TEM imaging were prepared using conventional mechanical polishing techniques, followed by Ar ion etching at 5 keV. Subsequently, a final Ar ion etching step at 2 keV was carried out to remove any surface damage caused during the previous steps. The TEM investigations were performed using an FEI Tecnai G2 TF 20 UT field emission TEM, operated at 200 keV to achieve a point resolution of 0.19 nm.